\documentclass[runningheads]{llncs}
\usepackage{graphicx,multirow,amssymb,booktabs,color}
\usepackage[colorlinks=true,linkcolor=black,citecolor=black,urlcolor=green,]{hyperref}
\usepackage[doipre={doi:~}]{uri}
\usepackage{marvosym}

\begin{document}
\title{X-Net: Brain Stroke Lesion Segmentation Based on Depthwise Separable Convolution and Long-range Dependencies}
\titlerunning{X-Net}
%

\renewcommand{\thefootnote}{\fnsymbol{footnote}} 

\author{Kehan Qi \inst{*1,2} \and
	Hao Yang  \inst{*1,2} \and
	Cheng Li \inst{1} \and
	Zaiyi Liu \inst{3} \and
	Meiyun Wang \inst{4} \and
	Qiegen Liu \inst{5} \and
	Shanshan Wang \inst{1}\textsuperscript{(\Letter)}}

\footnotetext{* These authors contruibuted equally to this work.}

\authorrunning{K. Qi et al.}
%
\institute{Paul C. Lauterbur Research Center for Biomedical Imaging, Shenzhen Institutes of Advanced Technology, Chinese Academy of Sciences, Shenzhen, Guangdong, China \\ 
	\email{sophiasswang@hotmail.com}
	\and
	University of Chinese Academy of Sciences, Beijing, China
	\and
	Department of Radiology, Guangdong General Hospital, Guangdong Academy of Medical Sciences, Guangzhou, Guangdong, China 
	\and
	Department of Radiology, Henan Provincial People's Hospital, Zhengzhou, Henan, China
	\and
	Department of Electronic Information Engineering, Nanchang University, Nanchang, Jiangxi, China  }
\maketitle              
\begin{abstract}
The morbidity of brain stroke increased rapidly in the past few years. To help specialists in lesion measurements and treatment planning, automatic segmentation methods are critically required for clinical practices. Recently, approaches based on deep learning and methods for contextual information extraction have served in many image segmentation tasks. However, their performances are limited due to the insufficient training of a large number of parameters, which sometimes fail in capturing long-range dependencies. To address these issues, we propose a depthwise separable convolution based X-Net that designs a nonlocal operation namely Feature Similarity Module (FSM) to capture long-range dependencies. The adopted depthwise convolution allows to reduce the network size, while the developed FSM provides a more effective, dense contextual information extraction and thus facilitates better segmentation. The effectiveness of X-Net was evaluated on an open dataset Anatomical Tracings of Lesions After Stroke (ATLAS) with encouraging performance achieved compared to other six state-of-the-art approaches.  We make our code available at \url{https://github.com/Andrewsher/X-Net}.
 
\end{abstract}

\section{Introduction}
Stroke causes the interruption of blood supply, and it is the second leading cause of death around the world \cite{[1]}. High-resolution brain MR images help specialists measure the stroke lesions and make effective treatment plans. Currently, the lesions are generally segmented manually by professional radiologists on MR images slice-by-slice, which is time-consuming and relies heavily on subjective perceptions. Therefore, automatic methods for brain stroke lesion segmentation are in urgent demand to get stroke measurements in the clinical practice. Nevertheless, this task is still challenging. First, the shape, scale, size, and location of lesions vary, which limits accurate auto-matic segmentation. Second, some lesions have fuzzy boundaries, confusing the confidential partition between stroke and non-stroke regions.

In the past few years, deep learning methods such as convolutional neural networks have achieved great success in the image segmentation task \cite{[3],[24]}. For example, SegNet \cite{[4]}, U-Net \cite{[5]} and 2D Dense-UNet \cite{[6]} are proposed based on symmetrical encoder-decoder architectures for image segmentation task. In addition, the dilated convolution operation \cite{[21]} and pyramid pooling architecture \cite{[22]} are introduced to obtain multi-scale feature maps and make reliable predictions. However, the application of these approaches is limited by the heavy network parameters. Furthermore, many automatic segmenting methods ignore the different sizes and locations of lesions, which is usually considered by experienced specialists according to a multi-scale context. This issue occurs since most current methods have not fully utilized contextual information among all the pixels. To address this issue, long short-term memory (LSTM) based networks \cite{[9]} are proposed to capture complex spatial contextual information, whose effectiveness relies heavily on long-term memorization. Furthermore, the Atrous convolution-based models \cite{[10],[11],[12]} are proposed to capture abundant multi-scale contexture information. Unfortunately, these methods still collect information from a few surrounding pixels, and cannot capture long-range dependencies veritably.

To address the two challenges mentioned above, we propose an end-to-end system, named X-Net, where the number of trainable parameters is much smaller than the existing methods, and the long-range dependencies are effectively explored for brain stroke lesion segmentation. Considering the effectiveness of Depthwise Separable Convolution (DSC) in reducing convolution kernel parameters \cite{[13],[14]}, this paper replaces the classical U-Net convolution operation with DSC. Moreover, a Feature Similarity Module (FSM) is designed to capture the long-range spatial contextual information, which contributes to the segmentation of lesions with different shapes and scales. This module can be plugged into any fully convolutional neural networks. In summary, we have developed an automated segmentation model with the following contributions:

1.	We design a non-local operation FSM to explore dense context information for effective brain lesion segmentation through extracting long-range dependencies.

2.	An X-Net framework that integrates the depthwise separable convolution and FSM is proposed, which facilitates better segmentation results with reduced trainable parameters.

3.	Our method achieves better results compared to six state-of-the-art methods on the Anatomical Tracings of Lesions After Stroke (ATLAS) dataset \cite{[2]}, which is an open-source dataset for the brain stroke lesion segmentation task.

\begin{figure}
	\includegraphics[width=\textwidth]{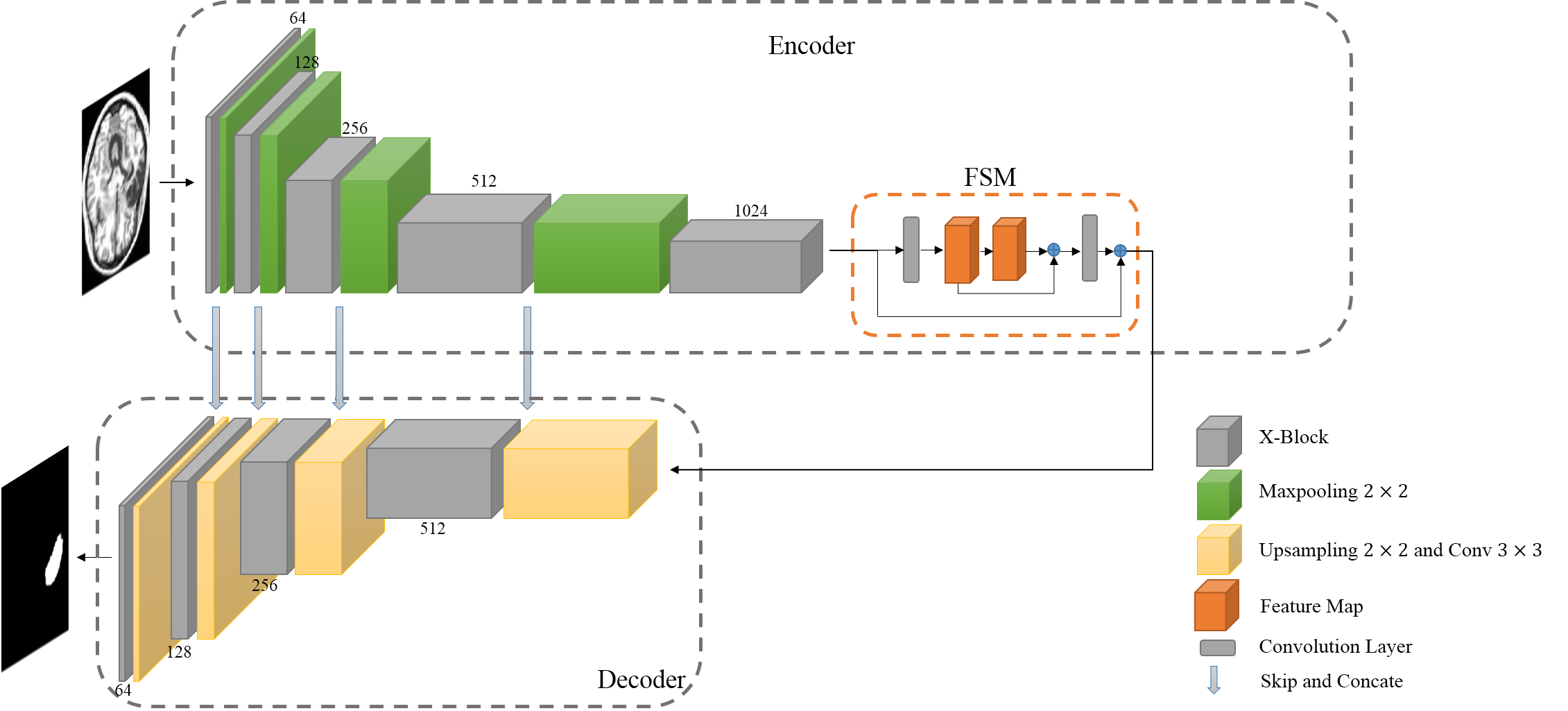}
	\caption{The illustration of the pipeline of our proposed method for brain stroke lesion segmentation. The numbers 64, 128, 256, 512 and 1024 indicate the number of filters.} \label{pipeline}
\end{figure}

\section{Method}
Fig.~\ref{pipeline} shows the pipeline of our proposed method for brain stroke lesion segmentation. We employ the encoder-decoder architecture and skip connections to improve segmenting performance, which has also been adopted in many segmentation tasks \cite{[3],[4],[5],[6],[7],[8]}. With the high-dimension features extracted by cascaded X-blocks, our proposed FSM efficiently calculates long-range dependencies through getting relations between any two positions in the feature map. A decoder architecture is then introduced subsequently to recover the spatial resolution.

\subsection{Feature Similarity Module for Long-Range Dependencies Extraction}
Dense context features for discrimination are essential in pixel-level visual tasks, which could be obtained by capturing long-range dependencies. In order to model abundant contextual relationships over feature representations, we propose a Feature Similarity Module (FSM). This module extracts a wide range of position-sensitive contextual information and encoded it into feature maps. Treating FSM as a network module that can be plugged to other fully convolutional neural networks, it may see wide applications in different situations for different tasks.

\begin{figure}
	\includegraphics[width=\textwidth]{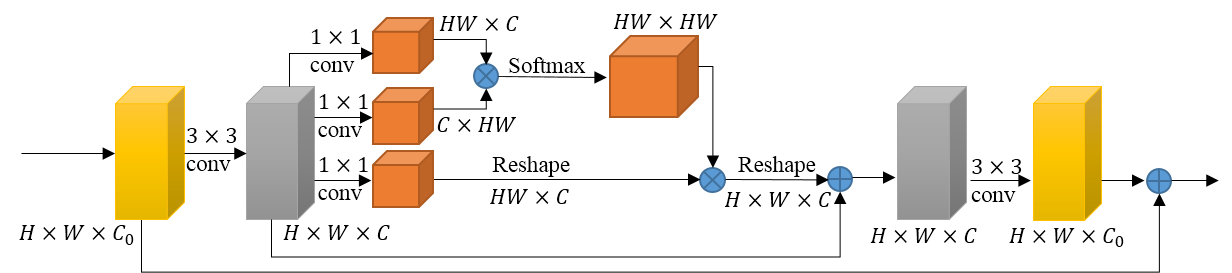}
	\caption{The details of Feature Similarity Module (FSM).} \label{FSM}
\end{figure}

As illustrated in Fig.~\ref{FSM}, given a feature map $X_0 \in R^{H\times W\times C_0}$, we first feed it into a convolution layer and generate a new feature map $X$ to filter out the irrelevant features, where $X \in R^{H\times W\times C}$ and $C < C_0$. In this work, we have $C = \frac{C_0}{8}$. For each pair of position $(x_i,x_j)$ in the feature matrix $X$, a relation map $f(x_i,x_j)\in R^{N\times N}$ is computed. Suggested by the non-local operation \cite{[15],[19]}, we define f as a combination of dot-product and softmax:
$$
f(x_i,x_j) = \frac {\exp \left( \alpha(x_i)^T \beta(x_j) \right)} {\sum\limits_{j=1}^N \exp \left( \alpha(x_i)^T \beta(x_j) \right)}
$$
where $f(x_i, x_j)$ measures the $j^{th}$ position’s impact on $i^{th}$ position, $\alpha(x_i)$ and $\beta(x_j)$ are embedded layers implemented by $1\times 1$ convolution, and $N$ is the number of positions in the feature map. It can be inferred that $f$ represents the relationships of all positions in the original feature map and captured dense contextual information. Meanwhile, we feed $X$ to a $1\times 1$ convolution layer to generate a new feature map $Y \in R^{H\times W\times C}$ which indicates the representation of the input signal and reshape it to $R^{N\times C}$. Furthermore, we multiply $f(x_i, x_j)$ by $Y$ and perform an element-wise sum operation with feature map $X$ as follows:
$$
Z_i = \sum\limits_{j=1}^N f(x_i, x_j)Y_j + X_i
$$

It can be inferred that the resulting feature map $Z$ is a sum of the relationship feature and the original feature. Therefore, $Z$ has a wide range of contextual view and effectively aggregates the long-range context. Finally, we feed $Z$ into a convolution layer to generate a feature map, which has the same shape with $X_0$, and perform a sum operation with $X_0$ as a residual block to avoid overfitting. 

\subsection{X-Net for Brain Stroke Lesion Segmentation}
U-Net \cite{[5]} uses original convolution for feature extraction, which may contain redundant structures. Thus, there is potential for further improvement. Moreover, the original U-Net model does not contain residual connection, which may lead to overfitting. Given the two considerations, we design a new basic block, X-block for our model. Specifically, depthwise separable convolution layer is employed to reduce the number of trainable parameters and ensure the strength of feature extraction and representation.

\begin{figure}
	\includegraphics[width=\textwidth]{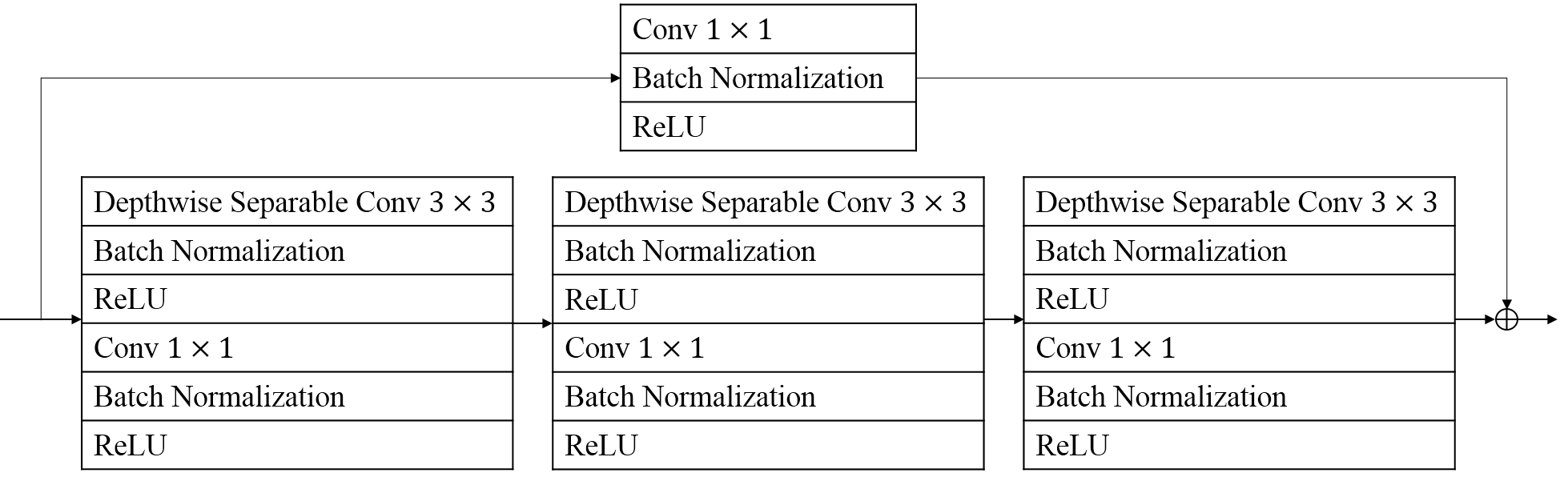}
	\caption{The details of X-block.} \label{x-block}
\end{figure}

A description of the X-block is given in Fig.~\ref{x-block}. A depthwise separable convolution is a convolution that is performed over each channel of the input feature map independently. Let $I\in R^{H\times W\times C_i}$ donates an input feature map of an X-block, where $C_i$ is the number of input channels. We feed $I$ into a depthwise separable convolution layer, which consists of a depthwise separable convolution followed by a $1 \times 1$ convolution, and generate a feature map $O \in R^{H\times W\times C_o}$, where $C_o$ is the number of output channels. We have 3 cascaded depthwise separable convolution layers in each of the X-block, and the kernel size of each separable convolution is $3\times 3$. For convenience, the residual connection consists of a $1 \times 1$ convolution layer to guarantee the number of output channels is $C_o$. It can be inferred that our X-block can largely reduce the number of parameters.

We design a neural network architecture named X-Net based on X-block and FSM (Fig.~\ref{pipeline}). The proposed segmentation model follows the encoder-decoder architecture and adopts the skip connections. X-blocks and max-pooling layers are cascaded in the encoder architecture to produce high-dimension feature maps, and FSM is employed to capture abstractive contextual information through extracting long-range dependencies. The decoder architecture composed of X-blocks and up-sampling layers is designed to recover the spatial resolution. After each convolution layer, we employ batch normalization and Rectified Linear Unit (ReLU).

\section{Experimental Results}
\subsubsection{Dataset}
To evaluate the performance of the proposed method on brain stroke lesion segmentation, our method is trained and validated on an open-source dataset, Anatomical Tracings of Lesions After Stroke (ATLAS) \cite{[2]}. This dataset consists of 229 T1-weighted normalized 3D MR images with diverse lesions manually segmented by specialists and is collected from 11 cohorts worldwide. Each of the 3D images is composed of 189 slices, and the size of each slice is $233\times 197$. In turn, normalized ATLAS dataset contains 43281 2D slices.

\subsubsection{Evaluation Metrics}
We evaluate the models by 5-fold cross-validation experiments. We select a series of evaluation metrics to quantify the performance of the proposed model, including Dice score, Intersection over Union (IoU), precision and recall. We calculate the evaluating scores for each 3D image in the validation set and report the average values. 

\begin{table}[tp]
	\centering  
	\fontsize{10}{12}\selectfont  
	\caption{Ablation analysis on ATLAS dataset for Feature Similarity Module.}\label{fsm_ablation}
	\begin{tabular}{cccccc}
		\toprule
		\bf\ \ Base Model\ \ &\bf\ \ FSM\ \ &\bf\ \ \ Dice\ \ \ &\bf\ \ \ IoU\ \ \ &\bf\ \ precision\ \ &\bf\ \  recall \ \ \\
		\hline
		\multirow{2}*{U-Net \cite{[5]}}	& 				& 0.4606 & 0.3447 & 0.5993 &0.4449 \\
										& \checkmark 	& 0.4749 & 0.3578 & 0.5862 & 0.4710 \\
		\hline
		\multirow{2}*{ResUNet \cite{[16]}}	& 				& 0.4702 & 0.3549 & 0.5941 & 0.4537 \\
											& \checkmark 	& 0.4792 & 0.3626 & 0.5891 & 0.4779 \\
		\hline
		\multirow{2}*{The proposed}	& 				& 0.4865 & 0.3699 & 0.6078 & 0.4702 \\
									& \checkmark 	& 0.4867 & 0.3723 & 0.6000 & 0.4752 \\
		\bottomrule
	\end{tabular}
\end{table}

\begin{table}
	\centering  
	\fontsize{10}{12}\selectfont  
	\caption{Comparison of brain stroke segmentation results on ATLAS dataset.}\label{comparison}
	\begin{tabular}{cccccc}
		\toprule
		\bf\ \ Method\ \ &\bf\ \ \ Dice\ \ \ &\bf\ \ \ IoU\ \ \ &\bf\ \ precision\ \ &\bf\ \  recall \ \ &\bf  \# Parameters \\
		\hline
		ResUNet \cite{[16]} & 0.4702 & 0.3549 & 0.5941 & 0.4537 & 33.2M \\
		DeepLab v3+ \cite{[7]} & 0.4609 & 0.3458 & 0.5831 & 0.4491 & 41.3M \\
		2D Dense-UNet \cite{[6]} & 0.4741 & 0.3559 & 0.5613 & \bf 0.4875 & 50.0M \\
		PSPNet \cite{[8]} & 0.3571 & 0.2540 & 0.4769 & 0.3335 & 48.1M \\
		SegNet \cite{[4]} & 0.2767 & 0.1911 & 0.3938 & 0.2532 & 29.5M \\
		U-Net \cite{[5]} & 0.4606 & 0.3447 & 0.5994 & 0.4449 & 34.5M \\
		X-Net (ours) & \bf 0.4867 & \bf 0.3723 & \bf 0.6000 & 0.4752 & \bf 15.1M \\
		\bottomrule
	\end{tabular}
\end{table}

\subsubsection{Implementation}
The proposed model is implemented in Keras. We use the Adam \cite{[23]} method to optimize our model.  We use the strategy of reduce learning rate on plateau to reduce learning rate automatically, in which the learning rate is reduced by a constant factor when the performance metric plateaus on the validation set. The initial learning rate is set to 0.001. We select a sum of Dice loss and Cross Entropy loss as the loss function. The batch size for training is set to 8, and the maximum number of epochs is set to 100. The experiments utilize NVIDIA RTX 2080 Ti with 11 GB memory. To adapt the proposed model, all slices are cropped into size $224\times 192$. 

\subsubsection{Ablation Analysis of Feature Similarity Module}
We employ the FSM to capture long-range dependencies and obtain dense contextual information. To verify the effectiveness of this module, we conduct experiments with different base models. It could be clearly observed from Table~\ref{fsm_ablation} that employing FSM yields better performance in three evaluation metrics (Dice, IoU and recall) compared to the base model, which demonstrates that FSM can help the model achieve better results consistently. Although there is little decrease in precision, the importance of recall is much higher than precision for brain stroke segmentation tasks as we need to make sure that all the strokes can be detected. Therefore, it is worthwhile to get a higher recall at the cost of a slight decrease on precision. Furthermore, Table~\ref{fsm_ablation} suggests that, FSM is more effective in U-Net and ResUNet than in our X-Net, which indicates that some of the interdependencies might have already been captured with our proposed X-block.

\subsubsection{Comparison to State-of-the-art Methods}
\begin{figure}
	\includegraphics[width=\textwidth]{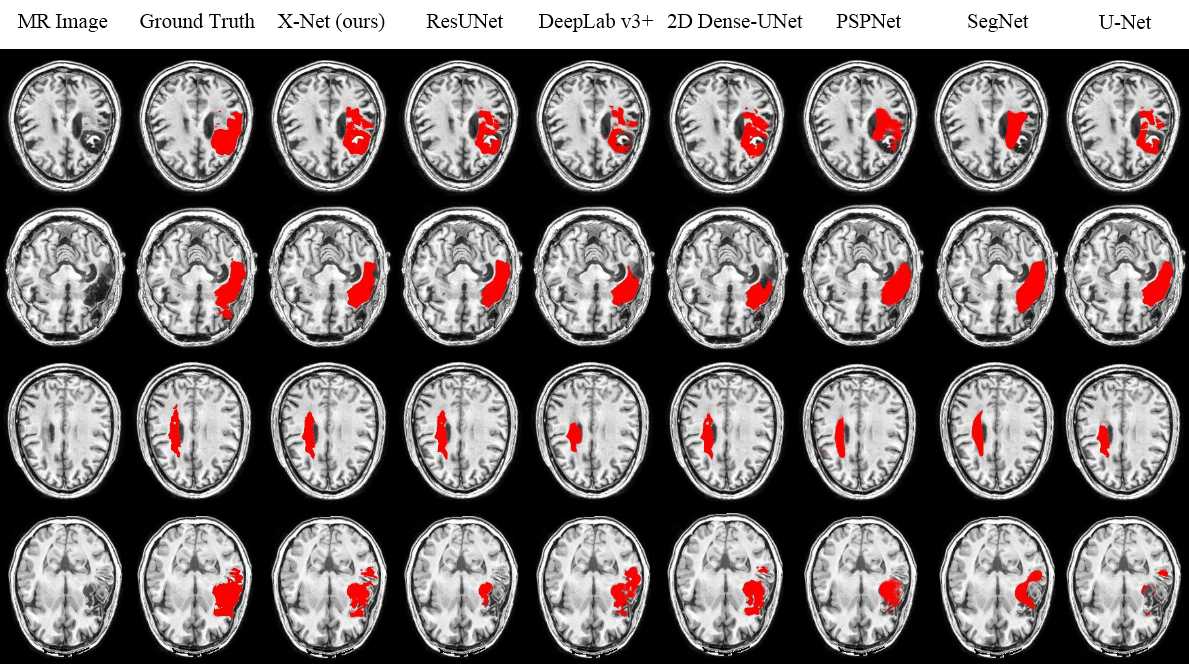}
	\caption{Examples of segmentation results on ATLAS dataset.} \label{pred_results}
\end{figure}

To validate the effectiveness of our proposed model, we compare our results to those of six state-of-the-art methods on the ATLAS dataset (Table~\ref{comparison}). It can be observed that our proposed model performs better than all the six methods with 0.0126, 0.0164 and 0.0006 improvement on Dice, IoU, and precision respectively. The experiment shows the good generalization capability and promising effectiveness of our proposed X-Net. Fig.~\ref{pred_results} shows some examples of the segmenting results. It can be inferred that our proposed X-Net can segment the brain stroke lesions in T1-weighted MR images very well. Furthermore, our X-Net has a significant smaller number of trainable parameters (15.1M), which could better fit the clinical requirements on fast image analysis.

\section{Conclusion}
We present an end-to-end model named X-Net for brain stroke lesion segmentation. X-Net can effectively extract informative features with fewer trainable parameters through the replacement of the traditional convolution with depthwise separable convolution. Furthermore, it can probe dense contextual information by the developed FSM. The proposed method gracefully addresses the problems of the existing approaches – the large number of parameters and the inefficiency in context capturing of long-range dependencies. Experiments on the ATLAS dataset demonstrates that our proposed X-Net could achieve better performance than existing models. 

\subsubsection{Acknowledgments} 
This research was partially supported by the National Natural Science Foundation of China (61601450, 61871371, 81830056), Science and Technology Planning Project of Guangdong Province (2017B020227012, 2018B010109009), Youth Innovation Promotion Association Program of Chinese Academy of Sciences (2019351), and the Basic Research Program of Shenzhen (JCYJ20180507182400762).

%
%
%
%

\end{document}